\documentclass[prb,twocolumn,floats,aps]{revtex4}
\usepackage{graphicx,graphics,color,epsfig,multirow} 
\usepackage{amsfonts}
\usepackage{amssymb}
\usepackage{physics}

\newcommand{\niu}{n_{i\uparrow}}
\newcommand{\nid}{n_{i\downarrow}}
\newcommand{\nis}{n_{i\sigma}}
\newcommand{\di}{\tilde d_i}
\newcommand{\mi}{\tilde m_i}
\newcommand{\dci}{\mathfrak d_i} 
\newcommand{\mci}{\mathfrak m_i} 
\newcommand{\Wc}{W_{\text{ch}}}
\newcommand{\Ws}{W_{\text{sp}}}
\newcommand{\xic}{\xi_{\text{ch}}}

\newcommand{\eps}{\epsilon}

\newcommand{\sig}{\sigma}
\newcommand{\uarw}{\uparrow}
\newcommand{\darw}{\downarrow}

\newcommand{\Ndc}{N_{\text{dc}}}

\begin{document}


\title{Charge and spin-specific local integrals of motion in a disordered Hubbard model}
%


\author{Brandon Leipner-Johns}
\author{Rachel Wortis}
\affiliation{Dept.\ of Physics \& Astronomy, Trent University, 1600 West Bank Drive, Peterborough, ON, K9L 0G2, Canada}


\date{\today}

\begin{abstract}
{\it Abstract}
While many-body localization has primarily been studied in systems with a single local degree of freedom, experimental studies of many-body localization in cold atom systems motivate exploration of the disordered Hubbard model.  With two coupled local degrees of freedom it is natural to ask how localization in spin relates to disorder in charge and vice versa.  
Most prior work has addressed disorder in only one of these sectors and often has not used measures of localization which distinguish between charge and spin.
Here we explore localization in the Hubbard model with a wide range of independent values of charge and spin disorder, 
using measures of localization based on charge and spin-specific integrals of motion.
Our results demonstrate symmetry between the response of the spin to charge disorder and vice versa, and we find very weak disorder in one channel, so long as the disorder in the other channel is sufficiently strong, results in localization in both channels.
The strength of disorder required in the less disordered channel declines as the system size increases.
Further, the weaker the disorder in the less-disordered channel, the longer the time scale at which localization appears in the dynamics of this degree of freedom.

\end{abstract}

\pacs{}

\maketitle




\section{Introduction}


How isolated quantum systems reach thermal equilibrium is a long standing question of continuing interest.\cite{Deutsch1991,Srednicki1994,Deutsch2018}
The absence of equilibration in some systems is also well known, notably Anderson localization in noninteracting systems with quenched disorder.\cite{Anderson1958}
The recent demonstration that the absence of equilibration can persist in the presence of interactions has launched the study of many-body localization (MBL).\cite{Gornyi2005,Basko2006,Oganesyan2007,Znidaric2008,Nandkishore2015,Imbrie2016,Abanin2018}

Most of the theoretical work in this area has focused on spin systems, or equivalently spinless fermions, in which there is just one local degree of freedom. 
However, interest in systems with multiple coupled degrees of freedom is growing because of the rich variety of new behaviour produced by their added complexity,\cite{Iadecola2019}
because of the question of how localization is affected by coupling to a bath\cite{Luitz2017,Hyatt2017},
and most directly because of significant experimental studies of MBL using cold atoms which are described by the Hubbard model.\cite{Schreiber2015,
Kondov2015,
Bordia2016,
Bordia2017}
Indeed, the influence of disorder on Hubbard systems has a long history of study given its relevance to the doping of high temperature superconductors and other transition metal oxides.
The Hubbard model contains two coupled local degrees of freedom, charge and spin.
A natural question to ask is how disorder in one of these effects the dynamics of each.

Most work to date has focused on charge disorder alone, by including randomly distributed site potentials in the model.
One study concluded that for sufficient charge disorder strength the average energy gap ratio is consistent with a Poisson distribution, characteristic of localization,\cite{Mondaini2015}
while another study of conserved quantities obtained through time averaging concluded instead that the system was neither localized nor generically ergodic.\cite{Mierzejewski2018}
Other studies have used dynamical properties to gain charge and spin-specific information suggesting that the charge is localized but the spin is neither localized nor generically ergodic.\cite{Prelovsek2016,Kozarzewski2018,Zakrzewski2018}
A very recent study goes further, arguing that the delocalized spin will cause the charge to also delocalize.\cite{Protopopov2018} 
There has also been a study on the case of spin disorder alone, adding a random magnetic field to the Hubbard model.\cite{Yu2018}  Focusing on the scaling of the entanglement entropy, the authors conclude similarly that the system is neither fully localized nor generically ergodic, but charge and spin-specific measures are not explored.
Two papers have examined a combination of charge and spin disorder, with equal strengths.\cite{Prelovsek2016,Mierzejewski2018}
Both conclude the system is fully localized with one\cite{Prelovsek2016} providing charge and spin-specific measures.
These measures used to gain information specific to charge and spin have been dynamical quantities.  
However, a number of authors have noted the potential for relevant time scales to be widely separated,\cite{BarLev2016prb,
Zakrzewski2018,
Protopopov2018} 
resulting in debate over whether numerics have captured the full dynamics.\cite{Zakrzewski2018}

A number of questions emerge from this context.  
Broadly, what is the localization behavior across the full spectrum of charge and spin disorder strengths?
In particular, is the level of localization in a given sector--charge or spin--simply a function of the disorder strength in that sector, or is there communication between them?
To address this, are there alternatives to dynamical measures which nonetheless provide charge and spin-specific information on localization?

In this work we introduce a method for constructing local conserved quantities from the local charge and spin degrees of freedom. 
We then measure the level of localization of the charge and spin using two approaches -- one based on the local conserved quantities and the other on the dynamics -- finding qualitative agreement between them.
We conclude that, while disorder in both channels is needed to achieve full localization, very weak disorder in one channel can result in nearly equal localization in both channels so long as the disorder in the dominant channel is sufficiently strong.
The strength of disorder required in the less disordered channel declines as the system size increases.
In addition, we find symmetry between the spin response to charge disorder and vice versa.

In Section \ref{sec_mm} we describe the model we study and provide details on (i) the definition and optimization of charge and spin-specific integrals of motion and (ii) the localization measures built from these integrals of motion as well as the dynamical quantities
 calculated for comparison.  Section \ref{sec_res} presents our results, which are discussed further in Section \ref{sec_disc}.

\section{Model and method}
\label{sec_mm}

We study a one-dimensional Hubbard model with nearest-neighbour hopping and on-site interactions, including both charge and spin disorder. 
\begin{eqnarray}
H &=& - t_h \sum_{\expval{ij},\sig} \qty( c_{i\sig}^{\dag} c_{j\sig} + h.c.) 
	+ U \sum_i \niu \nid \nonumber \\
& & 	\hskip 0.5 in + \sum_i \eps_i d_i + \sum_i h_i m_i 
\end{eqnarray}
where $d_i \equiv \niu +\nid$ is the local charge density operator and $m_i \equiv \niu - \nid$ is the local magnetization operator.
The charge and spin disorders are generated by randomly choosing $\eps_i$ and $h_i$ from uniform distributions $[-\Wc, \Wc]$ and $[-\Ws, \Ws]$, respectively.
We set the hopping amplitude as the unit of energy $t_h = 1$ so time is measured in units of $\hbar/t_h$. 
We focus primarily on half-filling and total spin zero.
At $U=0$, the localization behavior and dynamics of charge and spin are identical.
All results shown are for $U=1$.  
$U=8$ (data not shown) shows similar results.

Many measures of many-body localization have been proposed and implemented, including level statistics,\cite{Oganesyan2007} logarithmic time dependence and area law scaling of entanglement entropy,\cite{Znidaric2008,Bardarson2012,Bauer2013} memory of initial conditions,\cite{Schreiber2015,Kondov2015} etc.
Fully many-body localized systems can be described in terms of a macroscopic number of local conserved quantities: local integrals of motion.\cite{Serbyn2013,Huse2014}
An advantage of building a measure of localization around integrals of motion (IOMs) is that they are conserved, avoiding complications associated with determining how long a time scale is sufficient in a dynamical calculation. 
Meanwhile an advantage of building a measure around dynamical properties is the closer connection with experiments. 
Here we examine measures of both types.

\subsection{Charge and spin-specific integrals of motion}

A number of methods have been developed for identifying approximate local IOMs in large systems for which a full set of eigenstates is not known.\cite{Serbyn2013,Chandran2015,Ros2015,Rademaker2016, Rademaker2017,Inglis2016}
However, when all eigenstates are known, there is a very simple approach.\cite{Chandran2015,Rademaker2017,Wortis2017,Kulshreshtha2018,He2018,Goihl2018,Peng2019} 
Let $Q$ be the operator which generates the unitary transformation between the basis of local product states (the Fock basis), $\{\ket n\}$, and the basis of energy eigenstates,  $\{\ket{E_n}\}$:  $Q\ket n = \ket{E_n}$.
If $O$ is an operator that is diagonal in the Fock basis, then $QOQ^{\dag}$ is diagonal in the energy basis and hence commutes with the Hamiltonian, making it a conserved quantity, i.e.\ an integral of motion.\footnote{Note that this construction of the conserved operator is distinct from a unitary transformation as the operation is not applied to all states and operators.}
Specifically, if one chooses a local operator such as the number operator $\nis$, and if 
$Q$ also acts in a local way, then the resulting integral of motion $Q\nis Q^\dag$ can be argued to be local.\cite{Abanin2018}

Our interest is in examining the localization of the charge and spin degrees of freedom {\em separately}, and we therefore start with operators which are not only local but also charge and spin specific.  
$\di \equiv \frac{1}{\sqrt 2}(\tilde n_{i\uarw} + \tilde n_{i\darw})$ and $\mi \equiv \frac{1}{\sqrt 2}(\tilde n_{i\uarw}- \tilde n_{i\darw})$ are orthonormalized versions of the local charge density and magnetization operators, respectively, where $\tilde n_{i\sig} \equiv 2c_{i\sig}^\dag c_{i\sig} - I$.
Thus $(\di, \tilde d_j) = (\mi, \tilde m_j) = \delta_{ij}$ and $(\di, \tilde m_j) = 0$, where $(A,B) \equiv \frac{1}{N} \Tr(A^\dag B)$ is the Frobenius inner product, and $N$ is the number of states being traced over.
From these we construct conserved (and orthonormal) operators $\dci$ and $\mci$:
\begin{eqnarray}
\dci\ \equiv \ Q \di Q^{\dag} \ \ {\rm and} \ \ 
\mci \ \equiv \ Q \mi Q^{\dag}.
\end{eqnarray}

The unitary operator $Q$ is not unique, since any of the $N!$ one-to-one matchings $\ket n \leftrightarrow \ket{E_n}$ will also diagonalize the Hamiltonian. 
Which match is optimal?
A search through all relevant matchings to find the one that minimizes a chosen localization length is only possible in exceedingly small systems.\cite{Wortis2017}
Here we choose the matching that maximizes the weight of $Q$ on the identity, and therefore maximizes $\Tr Q$, similar to the approach taken by Ref.\ [\onlinecite{He2018}]. 
Computationally, this is identical to the well-known combinatorial optimization task known as the Assignment Problem, and we implement the Hungarian algorithm,\cite{Kuhn1955, Munkres1957,Papadimitriou1982} described in Appendix \ref{HA}.
A key feature of this method for our purposes is that it is unbiased towards the charge or spin sectors.

\subsection{Measures of localization}

To measure localization, we consider properties of both our IOMs and the dynamics of the system.
From the IOMs, we calculate three quantities: the single-site overlap, the overlap as a function of distance, and the overlap localization length. 
The single-site overlap is defined as the weight of an IOM on the local operator from which it was generated. For example, considering the IOM $\dci$, its single-site overlap is $(\di, \dci)$.
In practice, we average this quantity over the $L$ sites in the system and $\Ndc$ disorder configurations:
\begin{eqnarray}
O_c &\equiv& {1\over \Ndc}\sum_{\rm config} {1 \over L} \sum_i {1 \over N} \Tr(\di \dci) \\
O_s &\equiv& {1\over \Ndc}\sum_{\rm config} {1 \over L} \sum_i {1 \over N} \Tr(\mi \mci)
\end{eqnarray}
where the trace is performed over the half-filled, zero-net-spin subspace, and $N$ is the dimension of this subspace. 
Using a subspace with a fixed density of spin up and down fermions allows a clear investigation of system-size dependence, and omits single-particle states which naturally have the same charge and spin localization.
In a system with localized charge degrees of freedom, $O_c$ should tend to a nonzero value in the thermodynamic limit. In contrast, if the charge degrees of freedom are delocalized, $O_c$ should vanish in the thermodynamic limit. Similarly for $O_s$.

The overlap versus distance is defined in a similar manner, comparing the integral of motion at site $i$, referred to as the origin site, with the local density at a site a distance $\ell$ away. 
We average them over $\Ndc$ disorder configurations.
\begin{eqnarray}
O_i^c(\ell) &\equiv& {1 \over \Ndc} \sum_{\rm config} {1 \over N} \Tr(\tilde d_{i\pm \ell} \dci)\\
O_i^s(\ell) &\equiv& {1 \over \Ndc} \sum_{\rm config} {1 \over N} \Tr(\tilde m_{i\pm \ell} \mci) 
\end{eqnarray}
In this case, the trace is over the full Hilbert space in order to ensure the orthonormality $(\di, \tilde d_j) = (\mi, \tilde m_j) = \delta_{ij}$ indicating maximal overlap when $i=j$, or no overlap when $i\neq j$.
The terms $O_i^c(\ell)$ (for $\ell>0$) are then fit to an exponential $e^{-\ell/ \xic}$ to extract a charge localization length $\xic$, and similarly for spin.

These overlap measures allow us to distinguish between charge and spin localization and also share with many other measures used elsewhere a common foundation in the expectation values of local operators.

As for dynamical quantities, we quantify the memory of the initial charge and spin configurations using the local charge and spin correlations\cite{Prelovsek2016, Zakrzewski2018}
\begin{eqnarray}
D(t) &\equiv& D_0 \sum_i \bra \psi (d_i(t) - {\bar d})(d_i(0)-{\bar d}) \ket \psi \\
M(t) &\equiv& M_0 \sum_i \bra \psi m_i(t) m_i(0) \ket \psi
\end{eqnarray}
where $D_0$ and $M_0$ are chosen such that $D(0) = M(0) = 1$. 
We average over many disorder configurations and initial product states $\ket \psi$ in the half-filled ($\bar d = 1$), zero-net-spin subspace.
Also plotted are the saturation values of the local charge and spin correlations $D(\infty)$ and $M(\infty)$ 

Finally, a note on boundary conditions:  When calculating the overlap versus distance, to maximize the distances accessible, we use open boundary conditions.  To check if this results in edge effects which bias our results, we have compared the results for different origin sites, for both directions from the origin site, and for periodic boundary conditions.  In all cases the results show a consistent rate of decay within uncertainties.  For the single-site overlap, however, for which distance is not an issue, we use periodic boundary conditions.  Further details are provided in Supplementary Material.\cite{SM}

\section{Results}
\label{sec_res}

\begin{figure}[htbp] 
   \centering
   \includegraphics[width=\columnwidth]{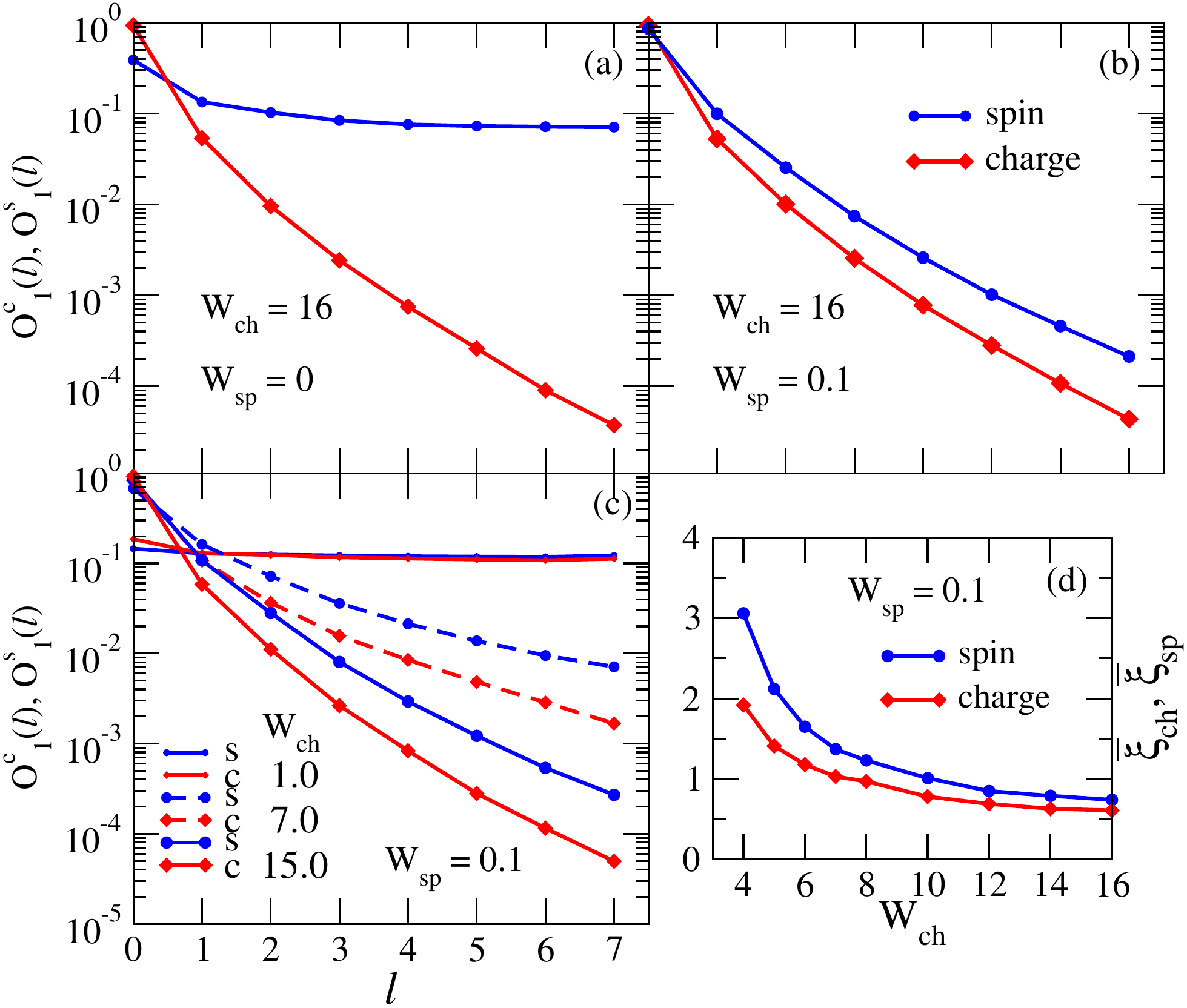} 
   \caption{Charge and spin overlap versus distance (a) with charge disorder alone and (b) with charge disorder plus a very weak spin disorder.  (c) Charge and spin overlap versus distance with weak spin disorder and three values of charge disorder.  
(d) Localization length, extracted from overlap versus distance, versus charge disorder strength for fixed weak spin disorder and $\Wc$ values for which the exponential fit has a correlation coefficient of 0.9 or higher.
Open boundary conditions, $L=8$, $U=1$, 1000-6000 disorder configurations.}
   \label{figWsp}
\end{figure}

Fig. 1 shows the dependence on distance of the charge and spin overlap.  With charge disorder alone, Fig. 1(a), the charge overlap decays exponentially with distance, indicating that the charge IOMs are spatially localized, but the spin overlap plateaus.  However, even a very small amount of disorder in the spin, Fig. 1(b), results in nearly the same localization in both charge and spin.  

To explore the evolution of this behavior with the strength of the charge disorder, Fig.\ \ref{figWsp}(c) shows the decay of the charge and spin overlaps with distance for a fixed weak value of spin disorder and three values of charge disorder.  
When the disorder in both charge and spin is small, there is no localization in either channel.
However, when the charge disorder is above a threshold, both charge and spin become localized, despite the spin disorder remaining very small.
Our results show exponential decay in both charge and spin for $\Wc \approx 7$ and above.
Fig.\ \ref{figWsp}(d) emphasizes this point by showing the localization lengths for charge and spin extracted from the overlap data at more values of the charge disorder strength.
Both charge and spin respond similarly to changes in the charge disorder alone.
The difference between the charge and spin response may become even smaller for larger systems, as discussed further below.

\begin{figure}[htbp]
\begin{center}
\begin{tabular}{cc}
(a) & (b) \\
\includegraphics[width=0.48\columnwidth]{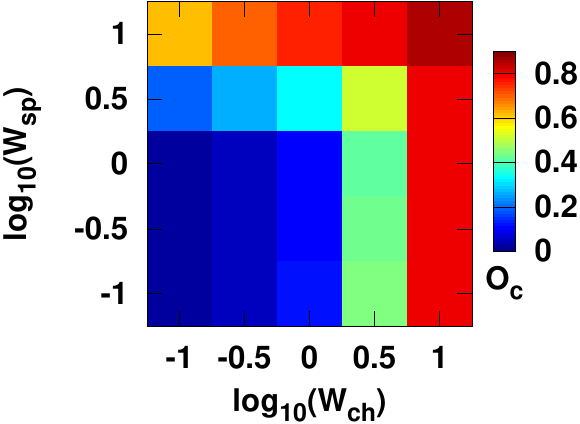}&
\includegraphics[width=0.48\columnwidth]{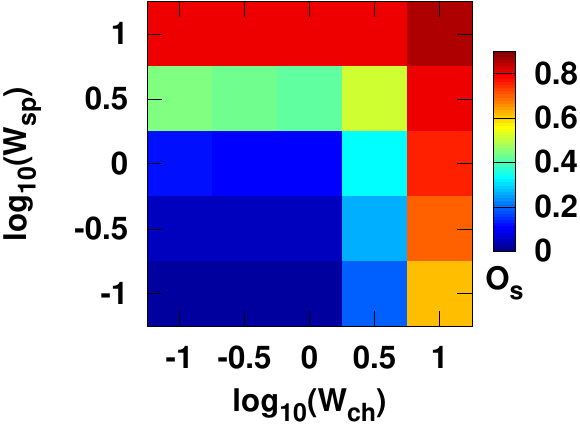}
\end{tabular}
\end{center}
\caption{Charge (a) and spin (b) single-site overlaps versus charge and spin disorder strength. Open boundary conditions, $L=6$, $U=1$, $10^4$ disorder configurations.}
\label{figsym}
\end{figure}

Thus far we have focused on situations in which the disorder in the charge dominates.  
In prior work, the case of charge disorder alone has received the most attention,\cite{Prelovsek2016,Mierzejewski2018,Kozarzewski2018,Zakrzewski2018,Protopopov2018} but spin disorder alone has also been studied.\cite{Yu2018}
In both of these limits, the channel which is disordered is generally seen to be localized while the other is not.
We have explored the full spectrum between these two limits, Fig.\ \ref{figsym}.
The localization of charge grows smoothly with the charge disorder strength largely independent of the spin disorder for $\Ws<\Wc$.
Meanwhile, spin, which is delocalized without spin disorder, shows a very sharp increase in localization, reaching localization comparable to that of the charge, at spin disorder strengths an order of magnitude less than that of the charge.

Note also the symmetry between charge and spin:  When Fig.\ \ref{figsym}(a) is reflected across the diagonal, it is the same as Fig.\ \ref{figsym}(b) to within the relative error.  This equivalence of the response of charge and spin to disorder in this model can be derived by applying a particle-hole transformation in just the spin-down component using the unitary operator $T = \prod_i (c_{i\downarrow}+c_{i\downarrow}^{\dag})$.  This transformation exchanges charge and spin ($T\di T^{\dag} =\mi$), and for a bipartite lattice and symmetric disorder distributions the Hamiltonian is mapped to another of the same form with the charge and spin disorder distributions switched.  See Appendix \ref{sym} for details.

\begin{figure}[htbp] 
   \centering
   \includegraphics[width=\columnwidth]{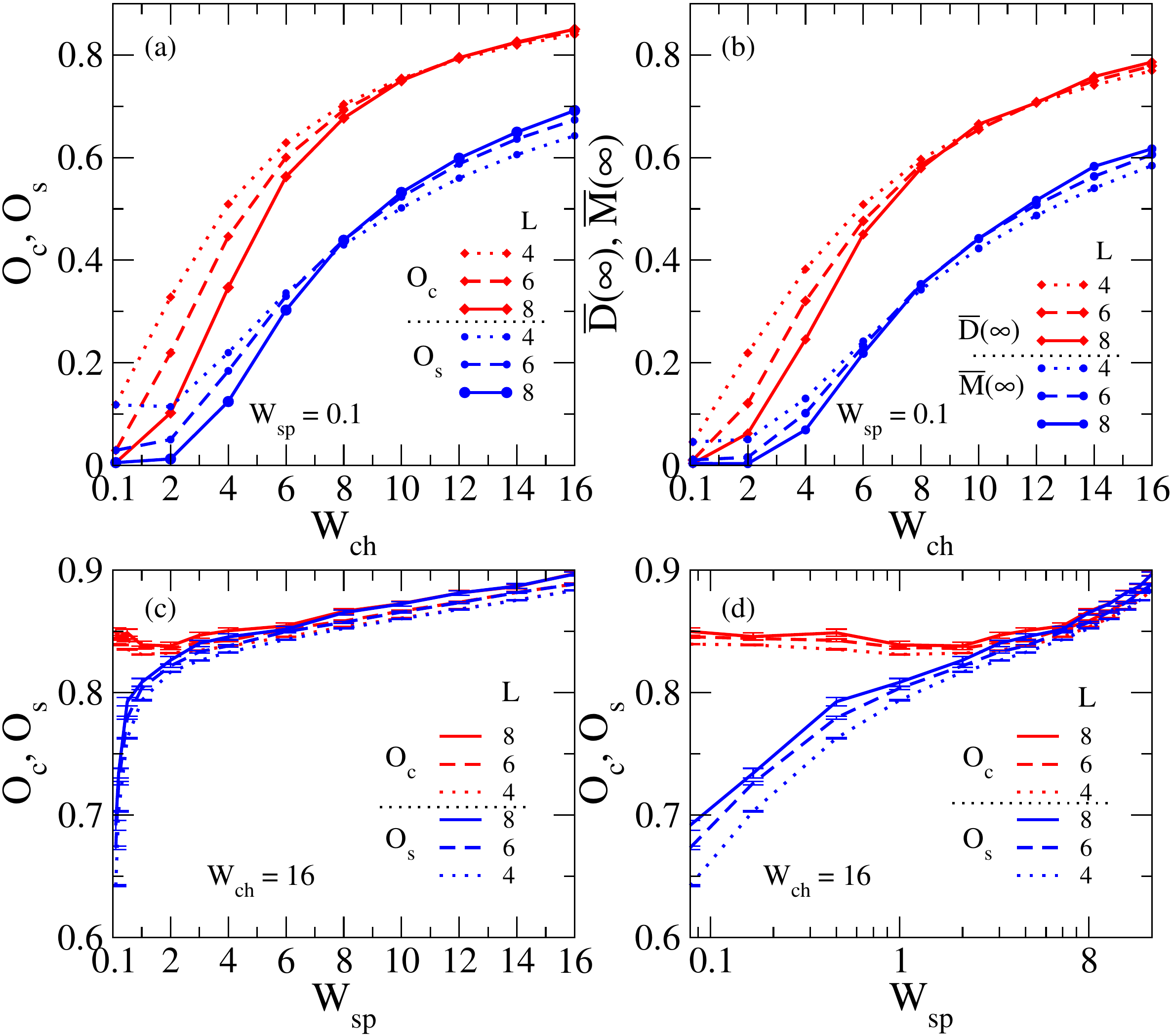} 
   \caption{Average charge and spin single-site overlap (a)
and saturation value of the local charge and spin correlations (b)
versus charge disorder for fixed weak spin disorder.  Charge and spin single-site overlap versus spin disorder for fixed strong charge disorder on linear (c) and log scale (d).
Periodic boundary conditions; $U=1$; $10^5$, $10^4$, and $10^3$ disorder configurations for $L=4$, 6, and 8, respectively.}  
   \label{figdyns}
\end{figure}

Because much of the work to date has focused on time-dependent quantities, we note that the parameter dependence of our IOM-based measures are consistent with that of the dynamics.
Fig.\ \ref{figdyns} compares the variation with charge disorder strength of the charge and spin single-site overlap (a) to that of the saturation value of the local charge and spin correlations (b).
The two figures are qualitatively the same and even very similar quantitatively.
In our small systems, only a broad crossover is visible, but the variation from $L=4$ to $L=8$ is suggestive of the expected evolution at larger system sizes to an abrupt transition.
Specifically, at high disorder the single-site spin overlap moves to larger values as the system size is increased, while at low disorder the overlap decreases as the system size is increased.\cite{Mondaini2015}
Fig.\ \ref{figdyns}(c) and (d) show the variation of the single-site overlap with $\Ws$ for $\Wc=16$.  Here, although the magnitude of the spin overlap drops quickly below $\Ws \sim 2$, at all $\Ws$ values the magnitude increases with increasing system size, suggesting that even for small values of spin disorder both degrees of freedom are localized (for sufficiently large $\Wc$).

\begin{figure}[htbp] 
   \centering
   \includegraphics[width=0.8\columnwidth]{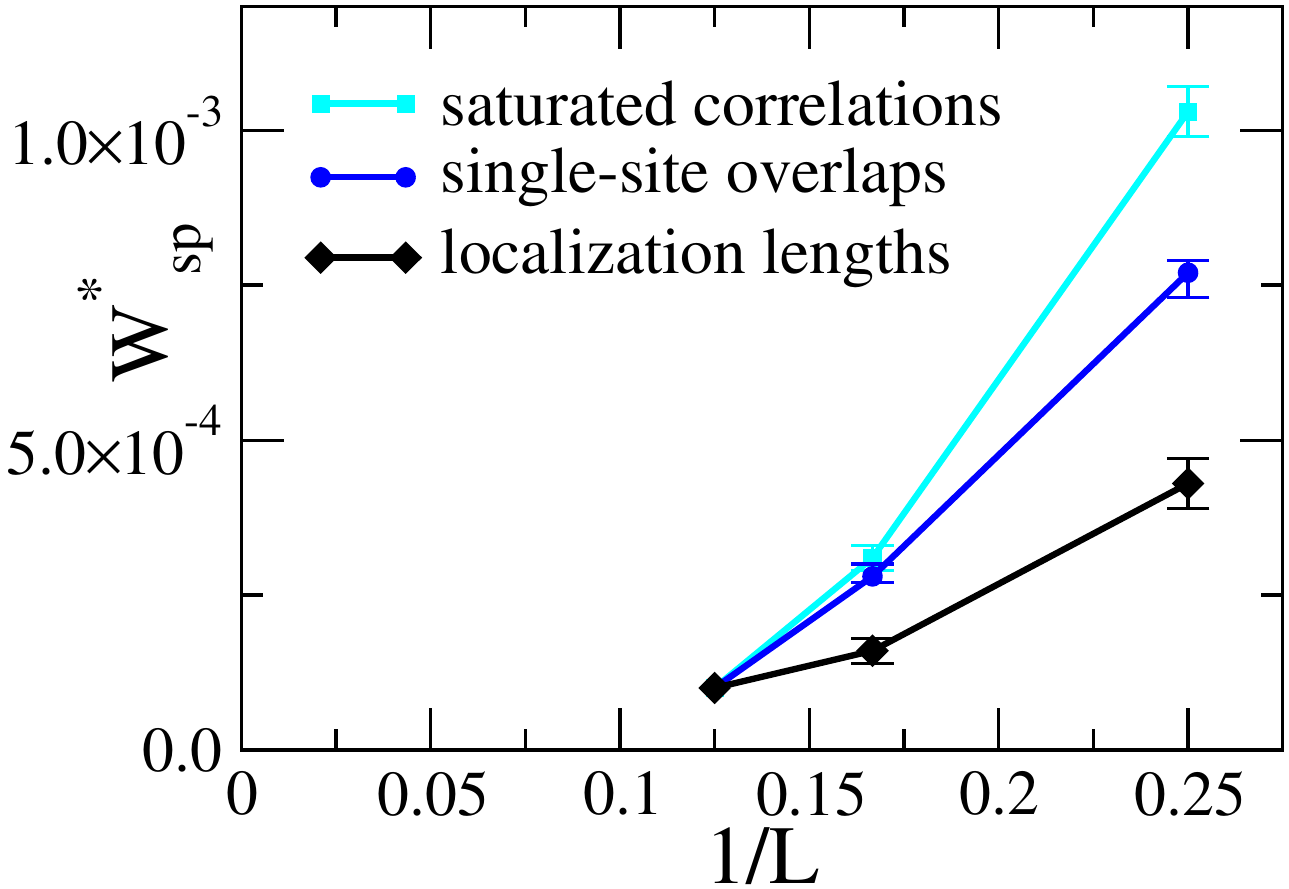} 
   \caption{Spin disorder strength required to obtain a fixed level of localization versus $1/L$ using three measures of spin localization:  localization length calculated from the overlap versus distance, single-site overlap, and the saturated correlation value.  Error bars indicate 2-3\% difference from the target value.
Data averaged over 2000-4000 disorder configurations for $L=8$, 1-5$\times 10^4$ for $L=6$, and 1-6$\times 10^5$ for $L=4$.}  
   \label{figsize}
\end{figure}

To explore further the behavior in the limit of vanishing spin disorder, Fig.\ \ref{figsize} shows as a function of $1/L$ the value of $\Ws$ necessary to obtain a fixed level of localization by three measures.  Starting from the spin localization length obtained with $\Ws=0.0001$ in an 8-site system, we searched for the value of $\Ws$ required to obtain the same localization length in a 6-site and a 4-site system.  
A similar analysis is also shown for the single-site overlap and for the saturated correlation values.  All cases are consistent with needing less spin disorder to obtain the same level of spin localization as the system size increases.  

\begin{figure}[htbp] 
   \centering
   \includegraphics[width=\columnwidth]{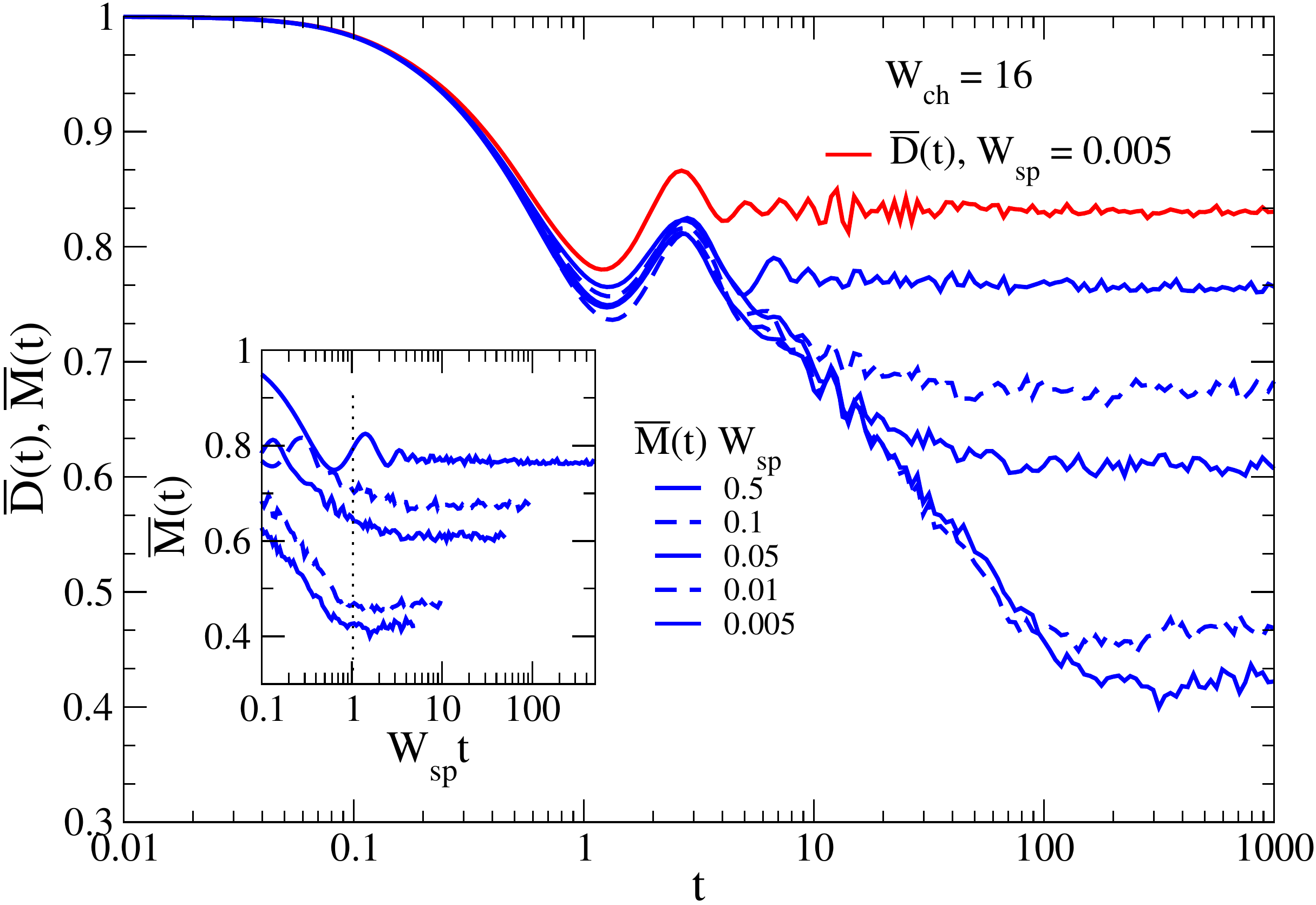} 
   \caption{Local charge and spin correlations versus time for fixed charge disorder $\Wc = 16$ and five different values of spin disorder.  Open boundary conditions, $L=8$, $U=1$, $10^3$ disorder configurations. 
   Inset shows the same spin correlation data versus $\Ws t$.}  
   \label{figdynt}
\end{figure}

A key issue in characterizing the system through its dynamics is the time scale at which localization will be visible.  
Fig.\ \ref{figdynt} shows the local charge and spin correlations versus time on a log scale for a fixed large value of charge disorder and a range of different spin disorder values.  
The turnover to the saturation value occurs at later times for smaller values of the spin disorder.  
The inset shows the same results with time measured in units of the inverse of the spin disorder, demonstrating that the localization is reflected in the dynamics at a time of $\sim 1/\Ws$.

\section{Discussion}
\label{sec_disc}


We have explored localization in the Hubbard model with a wide range of independent values of charge and spin disorder, 
using measures of localization which provide independent information on charge and spin.
We have focused on measures based on charge and spin-specific integrals of motion 
which are optimized for maximum locality.
Results for these measures are consistent with those based on dynamics, with the advantage that they avoid the question which arises in the case of time-dependent measures of whether sufficiently long times have been included.  

Our results show a symmetry between the response of the charge to spin disorder and vice versa.
We find that for sufficient disorder in one degree of freedom, only a small amount of disorder in the other degree of freedom localizes both degrees of freedom.
Indeed, the responses of charge and spin are clearly coupled in the sense that the level of localization in one sector is not simply dependent on the disorder in that sector.
These observations suggest that the reason for the delocalized behaviour observed in the case of charge disorder only\cite{Prelovsek2016,Mierzejewski2018} and spin disorder only\cite{Yu2018} is likely due to the presence of perfect symmetry in local spin and local charge respectively, as opposed to the absence of strong disorder in the second channel.
Consistent with this, we note that Ref.\ [\onlinecite{Mondaini2015}], which concluded that sufficient charge disorder did result in full localization, included a magnetic field at a single site in their system, breaking the local $SU(2)$ symmetry.  
Indeed, we have repeated this result, showing that a local field at a single site, in combination with strong charge disorder, is sufficient to allow localization to occur in both charge and spin, at sufficiently long times.
Interestingly Ref.\ [\onlinecite{Mondaini2015}] noted in the supplementary material that the effect of this single-site symmetry-breaking term increased with system size, similar to our results in Fig.\ \ref{figsize}.

We note that the time scale needed to observe localization in the dynamics of the less-disordered degree of freedom is proportional to the inverse of the disorder strength in that sector.
Thus, we suggest that in an experiment with a nearly uniform magnetic field it would take an exceedingly long time to observe localization in the spin degrees of freedom.
Indeed, it will be interesting in future work to explore further the time scale(s) associated with the coupling between the charge and spin degrees of freedom.  
Ref.\ [\onlinecite{Zakrzewski2018}] noted that even with charge disorder alone there was a crossover at long times to a slower decay of spin correlations, perhaps marking a time scale associated with coupling of the spin to the charge.  
The charge and spin-specific integrals of motion introduced here provide a convenient tool for this work.

In this work we have focused on half filling and $S_{tot}^z=0$, but varying these represents another avenue for future exploration.
Ref.\ [\onlinecite{BarLev2016prb}] considers disorder in the interaction term of the Hubbard model, resulting in different localization of singly and doubly occupied sites and hence significant filling dependence of the dynamics.
Ref.\ [\onlinecite{Yu2018}] also explores away from half filling, finding similarly distinct behavior at different fillings in the entanglement entropy.
In the absence of spin disorder, when $S_{tot}^z=0$, the local magnetization expectation values $\bra{E_n} m_i \ket{E_n}$ are confined to zero by symmetry.\cite{Prelovsek2016}  
In subspaces where $S_{tot}^z \ne 0$, this is no longer the case.
Our preliminary results for the variation with charge disorder of the distribution of these expectation values, in the absence of spin disorder, are suggestive of localization behavior.

\appendix

\section{Hungarian algorithm}
\label{HA}

Our goal is to find the arrangement of columns in a matrix which results in the greatest weight on the diagonal, i.e.\ maximizes $\sum_i|Q_{ii}|^2$, an example of a linear assignment problem.
While the naive approach would scale as $n!$ for an $n\times n$ matrix, the Hungarian algorithm, published in 1955 by Harold Kuhn, based on work by Hungarian mathematicians Konig and Egervary, originally scaled as $n^4$ and was later modified to scale as $n^3$.\cite{Kuhn1955, Munkres1957,Papadimitriou1982}
The algorithm is easiest to implement as a minimization procedure so we first convert to the matrix $M$ such that $M_{ij}=1-|Q_{ij}|^2$.
We then subtract from each column the value of its smallest element, thus ensuring that there is a zero in every column, and repeat the same process for each row.
If in an $n\times n$ matrix there are $n$ zeros which appear in every column and also in every row, the rearrangement of columns can then be done by inspection.  
However, this is not the case in general, and instead an iterative procedure alters the matrix (in a way which preserves the optimal solution(s)) until this form is achieved.
Efficiencies are gained from, for example, retaining information at each step on the locations of the minima, the location at which the search was stopped, etc.  Further details are available in Supplementary Material.\cite{SM}

\section{Spin-charge symmetry}
\label{sym}

Consider the unitary transformation of the Hamiltonian generated by the particle-hole operator $T = \prod_i (c_{i\downarrow}+c_{i\downarrow}^{\dag})$.
\begin{eqnarray} 
THT &=& -t \sum_{\langle ij\rangle \sigma} (c_{i\sigma}^{\dag} c_{j\sigma} + c_{j\sigma}^{\dag} c_{i\sigma}) 
\nonumber \\
& & + U\sum_i n_{i\uparrow} - U \sum_i n_{i\uparrow} n_{i\downarrow} 
\nonumber \\
& & + \sum_i \epsilon_i m_i + \sum_i h_i d_i + \sum_i \epsilon_i - \sum_i h_i \\ 
&\sim& -t \sum_{\langle ij\rangle \sigma} (c_{i\sigma}^{\dag} c_{j\sigma} + c_{j\sigma}^{\dag} c_{i\sigma}) + U \sum_i n_{i\uparrow} n_{i\downarrow} 
\nonumber \\
& & + \sum_i \epsilon_i m_i + \sum_i h_i d_i 
\label{line2}
\end{eqnarray}
Eq.\ (\ref{line2}) indicates the equivalence of the two Hamiltonians in the sense that they describe ensembles of systems with the same level of localization.  To see this consider the following points:  First, the last two terms create a uniform shift in energy with no change to the eigenstates.  Similarly for term two except the shift is specific to each block.  Next by changing the sign of creation operators on one sublattice of the bipartite lattice, the sign of the hopping term can be reversed.  Likewise, for symmetric disorder distributions switching the sign of the disorder terms will not affect disorder-averaged quantities.  Finally, reversing the sign of all terms will again not change the eigenstates.  Localization measures based on the full Hilbert space (or of a subspace for which the net spin and rescaled net filling are equal, $\expval{\sum_i \tilde m_i} = \expval{\sum_i \tilde d_i}$) will therefore show symmetry in spin and charge.

\begin{acknowledgments}
We gratefully acknowledge support by the Natural Sciences and Engineering Research Council (NSERC) of Canada.  
R.W. warmly thanks Malcolm Kennett for helpful discussions.
We are grateful to the referees for their constructive feedback.
\end{acknowledgments}


\onecolumngrid

\newpage

\begin{center}
{\bf \Large Supplementary Material}  \\
for \\
{\it Charge and spin-specific local integrals of motion in a disordered Hubbard model}
\end{center}
\vskip 0.2 in



\noindent
{\bf \large 1. Hungarian algorithm}
\vskip 0.2 in

This section describes the algorithm for choosing the ordering of columns in the matrix $Q$ which defines the optimized integrals of motion. [Kuhn 1955, Munkres 1957, Papadimitriou 1982]
The goal is to find the arrangement of the columns of $Q$ which results in the greatest weight on the diagonal, i.e. maximizes $\sum_i|Q_{ii}|^2$.
The algorithm is easiest to implement as a minimization procedure so we first convert to the matrix $M$ such that $M_{ij}=1-|Q_{ij}|^2$.
Note that $Q_{ij}\leq 1$ so $M_{ij} \geq 0$.
The process is summarized in Fig. 1 and described below.
\vskip 0.1 in

\begin{figure}[htbp] 
   \centering
   \includegraphics[width=6in]{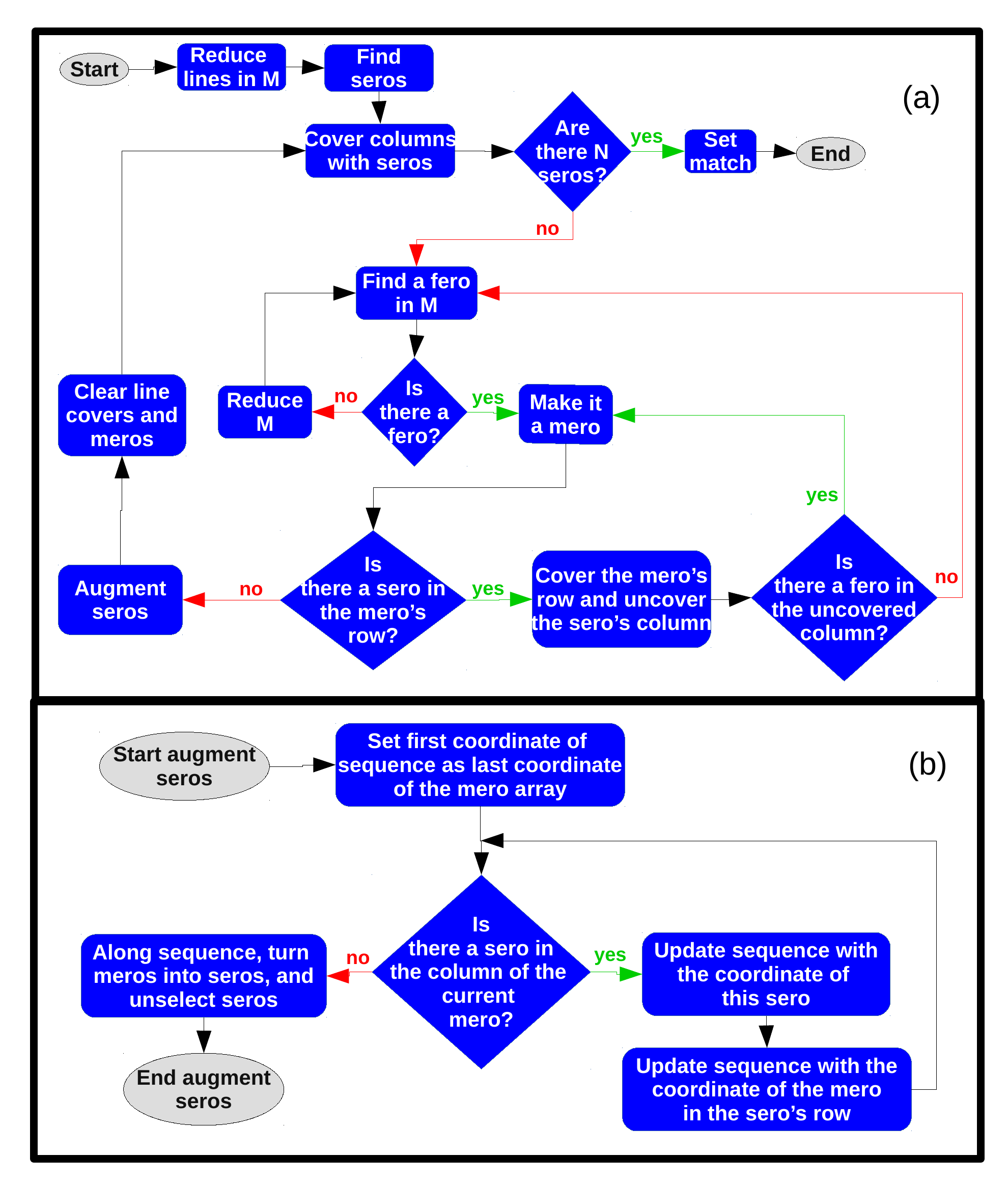} 
   \caption{Flow chart for implementation of Hungarian algorithm.  (a) Main process.  (b) Detail of `augment seros' step.}
   \label{fig:SMflow}
\end{figure}

\noindent
Description of Fig.\ \ref{fig:SMflow}(a): the main process: 
\vskip 0.1 in

\noindent
$\bullet$ {\bf Start.}  {\it Start algorithm with input $N \times N$ non-negative matrix $M$.} \\
$\bullet$ {\bf Reduce lines in $M$.} A line corresponds to either a row or column of the matrix. {\it Subtract the minimum element from every line.} This ensures that there is a zero in each line. \\
$\bullet$ {\bf Find seros.} A sero is a selected zero which shares neither a column nor a row with any other seros.  {\it Search the first column of $M$ for a zero.  This zero is selected and becomes a sero.  Record its coordinates.  Search the remaining columns in order, selecting no more than one zero per column and only zeros which do not share a row with previously selected zeros.}
Note that if there are $N$ seros, an optimal solution has been identified:  the columns can be rearranged to place each sero on the diagonal, minimizing the trace of $M$. \\
$\bullet$ {\bf Cover columns with seros.} {\it Cover each column that contains a sero.} This covering will ensure that later these columns are not searched. \\
$\bullet$ {\bf Are there $N$ seros?} {\it If yes, go to \textbf{set match}. If no, go to \textbf{find a fero in $M$.}} \\
$\bullet$ {\bf Set match.} {\it Set the matching by rearranging each column into the one corresponding to the row of its sero.} \\
$\bullet$ {\bf End.} {\it Stop, an optimal solution has been found.} \\
$\bullet$ {\bf Find a fero in $M$.} Feros are free zeros in the matrix, meaning they are not in covered rows or columns. {\it Search through all the uncovered elements of $M$ for zeros.} \\
$\bullet$ {\bf Is there a fero?} {\it If yes, go to \textbf{make it a mero.} If no, go to \textbf{reduce $M$.}} \\
$\bullet$ {\bf Make it a mero.} Meros are marked zeros in the matrix. We want to store their coordinates since they may become seros. {\it Store the coordinates of this element in the mero array.} \\
$\bullet$ {\bf Reduce $M$.} {\it Find the minimum uncovered element in $M$. Subtract it from each element of the uncovered columns, and add it to each each element of the covered rows.} This will create at least one fero. {\it Go back to \textbf{find a fero in $M$.}} \\
$\bullet$ {\bf Is there a sero in the mero's row?} {\it If yes, go to \textbf{cover the mero's row and uncover the sero's column}. If no, go to \textbf{augment seros.}} \\
$\bullet$ {\bf Cover the mero's row and uncover the sero's column.} This will ensure both the mero and sero are covered with only one line, allowing more possible zeros to be uncovered. \\
$\bullet$ {\bf Is there a fero in the uncovered column?} {\it If yes, go back to \textbf{make it a mero}. If no, go back to \textbf{find a fero in $M$.}} \\
$\bullet$ {\bf Augment seros.} See Fig.\ \ref{fig:SMflow}(b) and its description below. \\
$\bullet$ {\bf Clear line covers and meros.} {\it Uncover every row and column of the matrix.  
Clear the mero array. Go back to \textbf{cover columns with seros.}}
\vskip 0.1 in

\noindent
Description of Fig.\ \ref{fig:SMflow}(b): details of the `augment seros' step:
\vskip 0.1 in

\noindent
$\bullet$ {\bf Start augment seros.} {\it Start this algorithm with the mero array.}  This algorithm will build a sequence of coordinates with the first in the sequence being the coordinates of the last mero in the mero array, the one which did not have a sero in its row. \\
$\bullet$ {\bf Set first coordinate of sequence as the last in the mero array.} \\
$\bullet$ {\bf Is there a sero in the column of the current mero?} {\it If yes, go to \textbf{update sequence} with the coordinate of this sero. If no, go to \textbf{along sequence, turn meros into seros, and unselect seros}.} \\
$\bullet$ {\bf Update sequence with the coordinate of this sero.} \\
$\bullet$ {\bf Update sequence with the coordinate of the mero in the sero's row.} By construction, there must be a mero in this sero's row. {\it Set the next element of the sequence as the coordinate of this mero. Go back to \textbf{is there a sero in the column of the current mero.}} \\
$\bullet$ {\bf Along sequence, turn meros into seros, and unselect seros.} {\it Unselect every sero in the sequence. Select every mero in the sequence, converting them to seros.} Note that since the first and last elements of the sequence are meros, and the remaining elements alternate between meros and seros, the sequence initially contains one more mero than sero. Hence, after this step is applied, the total number of seros will increase by one. \\
$\bullet$ {\bf End augment seros.}


\newpage
 
\noindent
{\bf \large 2. Boundary condition data and discussion}
\vskip 0.2 in

Here we present a more detailed discussion of our choice to use open boundary conditions when studying overlap versus distance and periodic boundary conditions when studying the single-site overlap.
\vskip 0.1 in 

\noindent
{\bf A.  Overlap versus distance}
\vskip 0.1 in 

\begin{figure}[htbp] 
   \centering
   \includegraphics[width=5.5in]{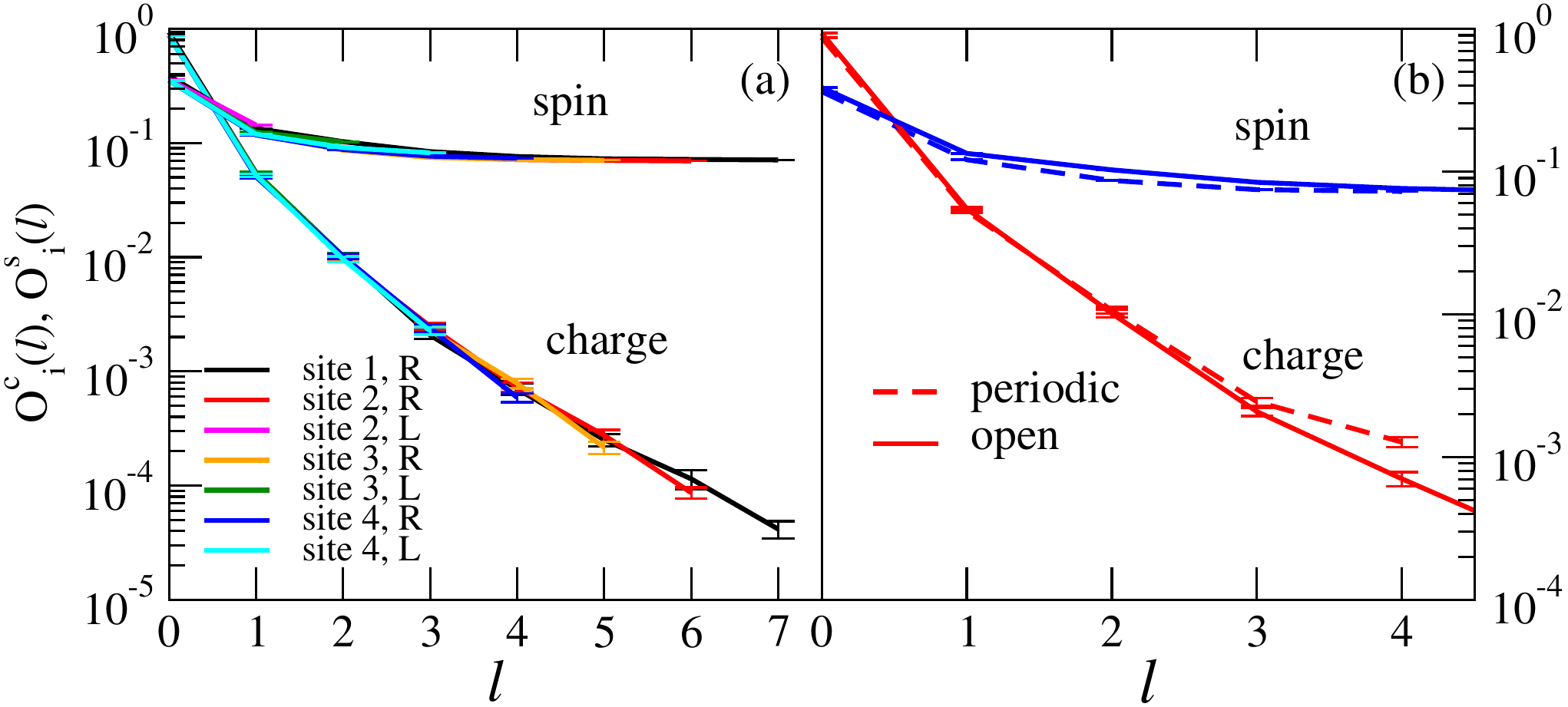} 
   \caption{Overlap versus distance.  
   (a) Open boundary conditions.  Results for multiple origin sites and for both directions.
   (b) Comparison between open and periodic boundary conditions.
   $L=8$, $U=1$, $\Wc=16$, $\Ws=0$, 4000 disorder configurations}
   \label{figSMbc}
\end{figure}

When we use overlap versus distance as a measure of localization, our focus is on the rate of decay.  Fig.\ \ref{figSMbc}(a) shows the overlap versus distance in an 8-site system for four origin sites:  $i=1,2,3,4$.  The figure shows separately the overlap going to the right and to the left of the origin site.  In all cases the decay is the same within uncertainties.  Fig.\ \ref{figSMbc}(b) compares the overlap versus distance obtained using periodic boundary conditions with that obtained using open boundary conditions.  The rate of decay out to $\ell=3$ is the same within uncertainties.  At $\ell=4$, the charge overlap with periodic boundary conditions is slightly higher than that with open boundary conditions.  This results from the overlap of the tails of the LIOM which extend in both directions from the origin site and hence overlap opposite the point of origin.
\vskip 0.1 in 

\noindent
{\bf B. Single-site overlap}
\vskip 0.1 in 

\begin{figure}[htbp] 
   \centering
   \includegraphics[width=3in]{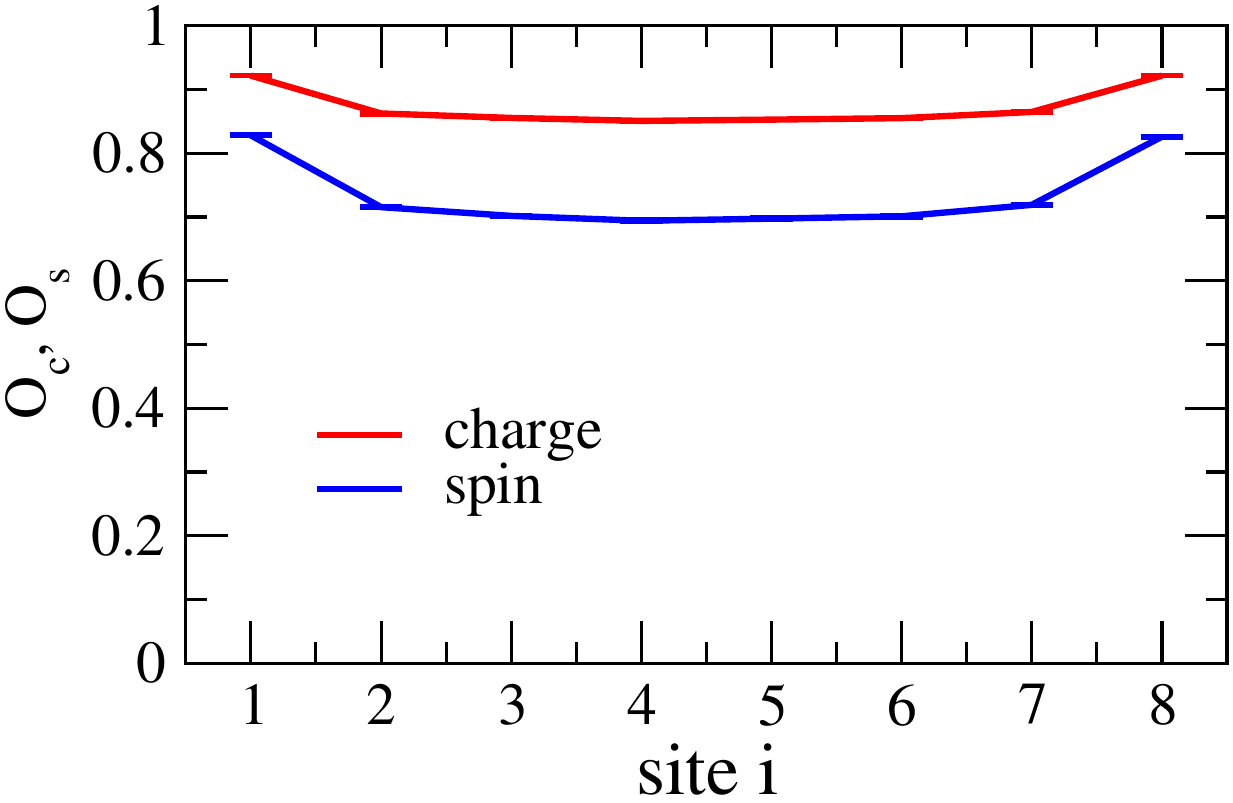} 
   \caption{Single-site overlap versus origin site.  Open boundary conditions.  $L=8$, $U=1$, $\Wc=16$, $\Ws=0.1$, 6000 disorder configurations}
   \label{figSMedge}
\end{figure}

In the case of the single-site overlap, access to large distances is not an advantage and sensitivity to edge effects is greater.  Both the single-site overlap itself and also the overlap versus distance at $\ell=0$ show some dependence on the origin site, with values at the edge being larger than those for interior sites.  These differences, while not distinguishable on the log scale used in our overlap versus distance graphs, are clearly visible in Fig.\ \ref{figSMedge}.  Their effect is mitigated by averaging over all origin sites.  Nonetheless, to ensure that small differences associated with edge effects do not influence our results, we use periodic boundary conditions when studying the system-size dependence using the single-site overlap.

\end{document}